\documentclass[12pt]{iopart}
\usepackage{graphicx}
\usepackage{color}
\begin{document}

\title[]{Electronic structure and magnetic properties of
the~spin-1/2 Heisenberg system CuSe$_2$O$_5$}

\author{O~Janson$^1$, W~Schnelle$^1$, M~Schmidt$^1$, 
Yu~Prots$^1$, S-L~Drechsler$^2$, S~K~Filatov$^3$, H~Rosner$^1$}
\address{$^1$Max-Planck-Institut f\"{u}r Chemische
Physik fester Stoffe, N\"{o}thnitzer Str.\ 40, 01187 Dresden, Germany}
\address{$^2$Leibniz-Institut f\"{u}r Festk\"{o}rper- und
Werkstoffforschung Dresden, P.O.\ Box 270116, 01171 Dresden,
Germany}
\address{$^3$Department of Crystallography, St. Petersburg State University, Universitetskaya nab.
7/9, St. Petersburg, 199034, Russia}

\ead{rosner@cpfs.mpg.de, janson@cpfs.mpg.de}

%\pagewiselinenumbers
%\renewcommand{\linenumberfont{\normalfont\bfseries\footnotesize}}

\begin{abstract}
A microscopic magnetic model for the spin-1/2 Heisenberg chain compound
CuSe$_2$O$_5$ is developed based on the results of a joint experimental and
theoretical study. Magnetic susceptibility and specific heat data give evidence
for quasi-1D magnetism with leading antiferromagnetic (AFM) couplings and an AFM
ordering temperature of 17~K. For microscopic insight, full-potential DFT
calculations within the local density approximation (LDA) were performed. Using
the resulting band structure, a consistent set of transfer integrals for an
effective one-band tight-binding model was obtained.  Electronic correlations
were treated on a mean-field level starting from LDA (\mbox{LSDA+$U$} method)
and on a model level (Hubbard model).  In excellent agreement of experiment and
theory, we find that only two couplings in CuSe$_2$O$_5$ are relevant: the
nearest-neighbour intra-chain interaction of 165~K and a non-frustrated
inter-chain coupling of 20~K. From a comparison with structurally related
systems (Sr$_2$Cu(PO$_4$)$_2$, Bi$_2$CuO$_4$), general implications for a
magnetic ordering in presence of inter-chain frustration are made. 
\end{abstract}
\maketitle

%\tableofcontents

\section{\label{intro}Introduction}
Low-dimensional spin-1/2 systems attract much interest due to a variety of
ground states (GS) found in these systems which originates from an interplay
between different exchange interactions and strong quantum fluctuations. There are, for
instance, the spin-Peierls GS in
CuGeO$_3$~\cite{CuGeO3_spin-Peierls_observation}, the helical GS in
LiCuVO$_4$~\cite{LiCuVO4}, and the quantum critical behaviour in
Li$_2$ZrCuO$_4$~\cite{Li2ZrCuO4,Li2ZrCuO4_new} etc. Besides, many of these materials (mostly
cuprates, vanadates and titanates) appeared to be realizations of theoretically
long-studied models in good approximation. One of the most prominent models is
the spin-1/2 nearest-neighbour (NN) chain described by the Heisenberg
Hamiltonian, for which the exact solution has been derived by
Bethe~\cite{Bethe_Ansatz}. The first compounds proposed to be good material
realizations of this model were Sr$_2$CuO$_3$ and
Ca$_2$CuO$_3$~\cite{Sr_Ca2CuO3,chi_Sr2CuO3}.  Recently, Sr$_2$Cu(PO$_4$)$_2$ and
Ba$_2$Cu(PO$_4$)$_2$ were suggested as even better
realizations~\cite{Sr2CuPO42}\nocite{chi_sr2cupo42,salunke,salunke_comment}--\cite{salunke_reply},
followed by a study of K$_2$CuP$_2$O$_7$~\cite{K2CuP2O7} that qualified this
compound to be the best realization of the spin-1/2 NN Heisenberg chain up to
date. As a natural consequence of its simplicity, this model poorly describes
one-dimensional and quasi-one-dimensional systems where additional interactions,
like longer range couplings or anisotropies, are present.  Thus, extensions of
this model are required to allow an accurate description of real materials. 

The simplest extension of the model is the inclusion of a next-nearest-neighbour
(NNN) coupling $J_2$ leading to the so-called zig-zag chain model. In case of an
antiferromagnetic (AFM) $J_2$ both NN and NNN couplings cannot be simultaneously
satisfied, in other words, the system is magnetically frustrated. Here, the
intra-chain frustration enriches the phase diagram with the spiral GS, the
gapped AFM GS and a quantum critical point at
$J_2$/$J_1=-0.25$~\cite{j1j2_phase_diag}\nocite{j1j2_chain_0.2411,j1j2_afm_region}--\cite{j1j2_fm_region}.
The evaluation of the two parameters in the zig-zag chain model allows to
estimate quantities which can be directly measured or derived from experiments,
namely spin-spin correlation functions, thermodynamic properties and the
response in high magnetic fields. Nevertheless, this model fails to describe
phenomena like long-range magnetic ordering, since one- or two-dimensional
systems do not order at finite temperatures according to the Mermin-Wagner
theorem. Thus, to account for magnetic ordering, the inter-chain (IC) coupling
has to be included in the model. This problem has been addressed in a series of
theoretical
works~\cite{tN_scalapino}\nocite{tN_schulz,essler97,tN_irkhin,tN_bocquet,tN_yasuda,tN_todo}--\cite{tN_zvyagin},
but the simplifications that had to be made to keep the models solvable (at
least approximately) impede an accurate description of complex situations. The
spin-1/2 Heisenberg chain system Sr$_2$Cu(PO$_4$)$_2$ is the most prominent
example for a huge discrepancy (two orders of magnitude) between the theoretical
(adopting a simple NN inter-chain coupling) and the experimentally observed
ordering temperature.  The origin of this discrepancy is hidden in the effects
of magnetic frustration and anisotropy, and the disentanglement of the effects
is difficult. Thus, a reliable theoretical description of this phenomenon is
still missing.

A natural way towards a deeper understanding is the search for real material
realizations of ``easy'' models. For such systems, the experimental data can be
supplemented by reliable microscopic models.  This way, joint experimental and
theoretical studies can challenge and improve the existing theoretical
approaches.

Cu$^{2+}$ phosphates are up to date the best realizations of the NN spin-1/2
Heisenberg chain model~\cite{Sr2CuPO42,salunke,K2CuP2O7} and could have a great
potential for the discovery of further low-dimensional systems.  Unfortunately,
the experimental information about these materials is rather limited since they
are up to now available as powders only, although several attempts have been
made to grow single crystals required for advanced experimental studies.

In this paper, the structurally closely related compound, CuSe$_2$O$_5$, is
investigated. Since selenites are often susceptible to chemical transport, the
advantage of this material is the potential to grow large single crystals of
high quality. Previous studies on a powder sample~\cite{cuse_magn_prop} may hint
a 1D character of its magnetic properties, but the low-temperature data are
strongly affected by impurities (figure~3 in~\cite{cuse_magn_prop}).  Therefore,
to probe the 1D nature of the system, a new detailed study on high quality
samples with lower defect concentration is desirable.

Though the chain-like arrangement of CuO$_4$ squares in CuSe$_2$O$_5$ is
topologically similar to that in Cu$^{2+}$ phosphates, the geometry of magnetic
coupling paths between the structural chains is essentially different. Thus,
the role of magnetic frustration, which is ruled by the inter-chain coupling,
can be evaluated in a comparative study.

The paper is organized as follows. In section~\ref{method} we describe the
synthesis, sample characterization and experimental as well as theoretical
methods used in this work. In section~\ref{sec_str}, we discuss the crystal
structure of CuSe$_2$O$_5$ in comparison to related systems. Section~\ref{res}
reports the results of our measurements and theoretical calculations and
proposes an appropriate microscopic model. A brief summary and an outlook are
given in section~\ref{summ}.

%%%%%%%%%%%%%%%%%%%%%%%%%%%%%%%%%%%%%%%%%%%%%%%%%%%%%%%%%%%
%
% Method and sample characterization
%
%%%%%%%%%%%%%%%%%%%%%%%%%%%%%%%%%%%%%%%%%%%%%%%%%%%%%%%%%%%

\section{\label{method}Method and sample characterization}

Single crystals of CuSe$_2$O$_5$ were grown by chemical vapour transport
using TeCl$_4$ as a transport agent. Using a micro-crystalline
powder of CuSe$_2$O$_5$ (obtained from a mixture of CuO and
SeO$_2$ at 723~K) as a source, the transport experiments were carried
out in an endothermic reaction of T$_2$ (source) 653~K to T$_1$
(sink)~553 K.

The obtained crystals have a green colour and form strongly elongated (along
$[$001$]$) plates, which macroscopically look like needles. The typical length
of a needle is 5--10~mm and the width does not exceed 1~mm and for most
crystallites it is considerably smaller. The slight disorientation of plates
forming a needle intricates a precise X-ray diffraction measurement on single
crystals. Thus, the samples were characterized by X-ray powder diffraction and
energy-dispersive X-ray spectroscopy (EDXS) experiments. The lattice parameters
of the synthesized crystals are similar to those reported for CuSe$_2$O$_5$
(table~\ref{str_table}). The results of the EDXS analysis (Cu 32.78$\pm$0.31, Se
67.14$\pm$0.23) for 13 points (2 crystals) yield
Cu:Se~$\approx$~0.488$\pm$0.006, very close to the ideal ratio of 0.5.  Thus,
the obtained single crystals represent an almost pure CuSe$_2$O$_5$ phase. 

Magnetization was measured in a SQUID magnetometer (1.8~--~350~K) in magnetic
fields up to 1~T. Heat capacity (1.8~--~100~K) was determined by a
relaxation method up to $\mu_0H$~=~9~T.

DFT calculations were carried out using the full potential local orbital code
(FPLO) version 7.00-27~\cite{fplo}. The standard basis set and the Perdew-Wang
parameterization of the exchange-correlation potential were
used~\cite{perdew_wang}. Strong on-site Coulomb interaction in the Cu 3$d$
orbitals, insufficiently described in the LDA, was taken into account
independently (i) by mapping the LDA antibonding Cu---O $dp_{\sigma}$ bands onto
a tight-binding model ($\hat{H} = \sum_{i}\epsilon_{i} + 
\sum_{<ij>\sigma}t_{ij}(c^{\dagger}_{i,\sigma}c_{j,\sigma} + \textrm{H.c.})$)
and subsequently via a Hubbard model ($\hat{H} =
\sum_{i}\epsilon_{i} + \sum_{<ij>\sigma}t_{ij}(\hat{c}_{i,\sigma}\hat{c}^{\dagger}_{j,\sigma} +
\textrm{H.c.}) +
U_{\rm{eff}}\sum_{i}\hat{n}_i,_{\uparrow}\hat{n}_i,_{\downarrow}$) onto a
Heisenberg model ($\hat{H} = \sum_{<ij>}{J_{ij}\hat{\vec{S_i}}\hat{\vec{S_j}}}$)
(the procedure is well justified for spin excitations in the strongly correlated
limit ($U_{\rm{eff}}~{\gg}~t_{ij}$) at half-filling (${\langle}n_i{\rangle}=1$))
and (ii) by using the \mbox{LSDA+$U$} method~\cite{lsdu} ($U_d$~=~6.5~eV,
$J_d$~=~1~eV). For the LDA calculations, we used a $k$-mesh of 1296 $k$-points
(355 points in the irreducible wedge), for \mbox{LSDA+$U$} calculations of
supercells irreducible $k$-meshes of 226, 242, 147 and 126 $k$-points were used.
All $k$-meshes are well converged.

Quantum Monte-Carlo simulations have been
preformed on $N=1200$ sites clusters of $S=1/2$ spins (30 coupled chains of 40 sites
each) using the ALPS software package~\cite{alps}.

%%%%%%%%%%%%%%%%%%%%%%%%%%%%%%%%%%%%%%%%%%%%%%%%%%%%%%%%%%%
%
% Crystal structure and empirical magnetic models 
%
%%%%%%%%%%%%%%%%%%%%%%%%%%%%%%%%%%%%%%%%%%%%%%%%%%%%%%%%%%%

\section{\label{sec_str}Crystal structure and empirical magnetic
models}

Crystal structures of cuprates are often subdivided into four large groups
according to their dimensionality, which reflects how their elementary building
blocks ---  CuO$_4$ plaquettes (planar or distorted) --- are connected: they can
be isolated (zero-dimensional, 0D) or form chains (one-dimensional, 1D), layers
(two-dimensional, 2D) or frameworks (three-dimensional, 3D). Although it is true
for many systems that the magnetic dimensionality follows the dimensionality of
the crystal structure, the real situations are often more complex, especially
for 0D cases. There, the magnetic dimensionality is ruled by (i) the orientation
of neighbouring plaquettes, and (ii) the position of anion groups formed by
non-magnetic atoms that bridge the magnetic plaquettes. In most cases, the
connection between structural peculiarities and the appropriate magnetic model
cannot be accounted for by applying simple empirical rules  (for instance,
Goodenough-Kanamori-Anderson rules).

Therefore, an almost complete understanding of the macroscopic magnetic
behaviour for a certain system of this class can be achieved only basing on a
relevant microscopic model.  The latter can be constructed either by using
advanced experimental techniques (for instance, inelastic neutron scattering) or
theoretical (DFT) calculations.  Naturally, the most reliable approach is the
combination of such a theory and experiment.  Due to the complexity of such an
analysis, it has been accomplished only for a rather limited number of real
systems. 

\begin{figure}[htbp]
\includegraphics[width=16cm]{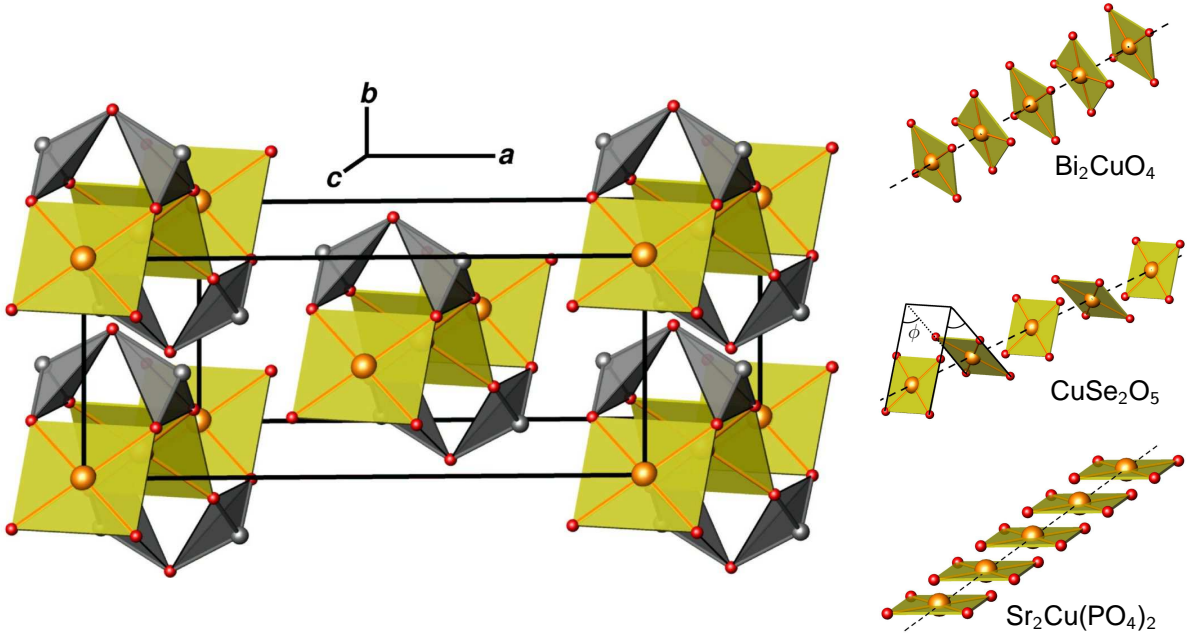}
\caption{\label{str}
Left panel: the crystal structure of CuSe$_2$O$_5$. Isolated CuO$_4$
plaquettes (yellow) are bridged by SeO$_3$ pyramids (gray) and form
chains running along $c$. The chains are closely stacked in $b$ 
direction, and well separated in the $a$ direction.  Right panel:
geometry of ``chains'' formed by isolated CuO$_4$ plaquettes.
Neighbouring plaquettes are stacked and twisted with respect to each
other in Bi$_2$CuO$_4$ (top), tilted in CuSe$_2$O$_5$ (middle) and form planar
edge-sharing chains with every second plaquette cut out in
Sr$_2$Cu(PO$_4$)$_2$ (bottom). The non-magnetic groups (BiO$_4$,
SeO$_3$ and PO$_4$ for Bi$_2$CuO$_4$, CuSe$_2$O$_5$ and
Sr$_2$Cu(PO$_4$)$_2$, respectively) bridging the neighbouring
plaquettes are not shown.}
\end{figure}

Two well studied systems of this class --- Bi$_2$CuO$_4$~\cite{Bi2CuO4} and
Sr$_2$Cu(PO$_4$)$_2$~\cite{Sr2CuPO42} --- are both structurally 0D cuprates, but
antipodes with respect to their magnetic behaviour. A drastic change of the
magnetic coupling regime originates from the arrangement of neighbouring
plaquettes  (figure~\ref{str}, right panel: top and bottom): stacking
(accompanied by additional twisting) of neighbouring plaquettes on top of each
other makes Bi$_2$CuO$_4$ a 3D magnet with
$T_{\rm{N}}$~$\approx$~47~K~\cite{bicu_tN} while in Sr$_2$Cu(PO$_4$)$_2$
the plaquettes are arranged in a planar fashion (formally reminiscent of an
edge-sharing chain with every second plaquette cut out), leading to a pronounced
1D behaviour and a very low N\'eel temperature
$T_{\rm{N}}$~=~0.085~K~\cite{chi_sr2cupo42}. In this context we mention that the
magnetic dimensionality of these systems is controlled by the dihedral angle
$\phi$ between neighbouring plaquettes (figure~\ref{str}, right panel, middle).
In CuSe$_2$O$_5$, the magnetic plaquettes are isolated (like in Bi$_2$CuO$_4$
and Sr$_2$Cu(PO$_4$)$_2$) but tilted with respect to each other forming a
dihedral angle $\phi$ of about 64$^{\circ}$, i.e. in between $\phi=0^{\circ}$
for the 3D Bi$_2$CuO$_4$ and $\phi=180^{\circ}$ for the 1D Sr$_2$Cu(PO$_4$)$_2$
(figure~\ref{str}, right panel). Another controlling parameter is the direct
Cu---Cu distance $d$.  Again, CuSe$_2$O$_5$ with $d$ close  to 4 \r{A} lies in
between $d=2.9$~\r{A} for Bi$_2$CuO$_4$ and $d=5.1$~\r{A} for
Sr$_2$Cu(PO$_4$)$_2$. Thus, CuSe$_2$O$_5$ is structurally in between the two
closely related systems: the 3D magnet Bi$_2$CuO$_4$ and the 1D magnet
SrCu$_2$(PO$_4$)$_2$. Does this analogy hold also for the magnetism? To answer
this question, additional arguments have to be addressed.

Besides $\phi$ and $d$, further structural features provide a deeper insight
into the crystal chemical aspects relevant for the magnetism.  In CuSe$_2$O$_5$,
two SeO$_3$ pyramids sharing an oxygen atom (forming Se$_2$O$_5$ polyanions)
bridge neighbouring CuO$_4$ plaquettes (see figure~\ref{str}). This structural
peculiarity is reflected in the morphology of the synthesized crystals
(section~\ref{method}): the needle-like shape with an elongation along $[$001$]$
fits perfectly to structural chains along $c$ (figure~\ref{str}) formed by
alternation of CuO$_4$ plaquettes and Se$_2$O$_5$ polyanion groups.  From the
topological similarity of this structural chain to the one in the structure of
Sr$_2$Cu(PO$_4$)$_2$ (there, neighbouring plaquettes are bridged by two PO$_4$
tetrahedra) a 1D behaviour of CuSe$_2$O$_5$ might be expected. A second argument
supports this proposition: in CuSe$_2$O$_5$, the structural chains are not
connected by covalent bonds, making a strong inter-chain coupling unlikely,
similar to Sr$_2$Cu(PO$_4$)$_2$, where the neighbouring chains are well
separated by Sr cations. These similarities of CuSe$_2$O$_5$ and
Sr$_2$Cu(PO$_4$)$_2$ may lead to the conclusion that both systems imply
essentially the same physics. However, a closer inspection of more subtle
crystal chemical aspects immediately reveals an important difference related to
the inter-chain coupling. As it follows from the microscopic
model~\cite{Sr2CuPO42}, in Sr$_2$Cu(PO$_4$)$_2$ there are two relevant NN
inter-chain couplings (2.7~K), which are equivalent by symmetry.  Together with
an intra-chain NN coupling (187~K), they induce magnetic frustration which
commonly leads to a considerable decrease of the ordering temperature
($T_{\rm{N}}$=0.085~K). In CuSe$_2$O$_5$, these two inter-chain couplings are
not symmetry-equivalent. Therefore, by reducing the strength of one of them, the
frustration can be lifted. 

So far, crystal chemical considerations provided us with a qualitative insight.
For a quantitative model, a microscopic analysis is required.  Thus, in the
next section a microscopic model  basing on the results of DFT calculations is
constructed.

A prerequisite for an accurate modeling based on a band structure code is reliable
structural information. For CuSe$_2$O$_5$, two refinements of the same
structural model (space group $C2/c$ with four formula units per cell) have been
proposed so far~\cite{str_old,str_new}.  Both structural data sets agree quite
well with each other and with the lattice parameters of the synthesized samples
(see table~\ref{str_table}).\footnote{It is well known, that X-ray diffraction
analyses may result in considerable inaccuracies for internal coordinates of
light elements (especially, hydrogen).  These inaccuracies can have a large
impact on the magnetic properties~\cite{kagome_kapel_hayd_PRL}. Since there are
no light atoms in CuSe$_2$O$_5$, we rely on the diffraction analysis and
therefore, no structural relaxation has been performed.} The reliability of the
structural data has been indirectly confirmed \textsl{a posteriori} by the good
agreement of calculated and experimentally measured quantities.

\begin{table}[htbp]
\caption{\label{str_table}Comparison of measured lattice parameters $a$, $b$,
$c$, the monoclinic angle $\beta$ and the unit cell volume $V$ of
CuSe$_2$O$_5$ with previously published data.}
\begin{indented}
\item[] \begin{tabular}{@{}lllll}
		\br parameter & Ref.~\cite{str_new} &
		Ref.~\cite{str_old} & this work \\
		$a$, \r{A} & 12.3869 & 12.254 & 12.272 \\
		$b$, \r{A} & 4.8699 & 4.858 & 4.856\\
		$c$, \r{A} & 7.9917 & 7.960 & 7.975 \\
		$\beta$, $^{\circ}$ & 109.53 & 110.70 & 110.91 \\
		$V$, \r{A}$^3$ & 447.13 & 443.27 & 443.95 \\ \br
	\end{tabular}
\end{indented}	
\end{table}

\section{\label{res}Results and discussion}

%%%%%%%%%%%%%%%%%%%%%%%%%%%%%%%%%%%%%%%%%%%%%%%%%%%%%%%%%%%
%
% Thermodynamical measurements 
%
%%%%%%%%%%%%%%%%%%%%%%%%%%%%%%%%%%%%%%%%%%%%%%%%%%%%%%%%%%%

\subsection{Thermodynamical measurements}

The first probe for the magnetic properties of a certain system is the
measurement of magnetization ($M$) at various temperatures in a constant field
($H$) yielding the temperature dependence of magnetic susceptibility
($\chi(T)=M(T)/H$).  This measurement already yields valuable information on the
magnetic dimensionality, the sign and the energy scale of leading couplings, the
presence of a spin gap, the spin anisotropy and the quality (defects, purity) of
a sample.

\begin{figure}[htbp]
\includegraphics[angle=270,width=16cm]{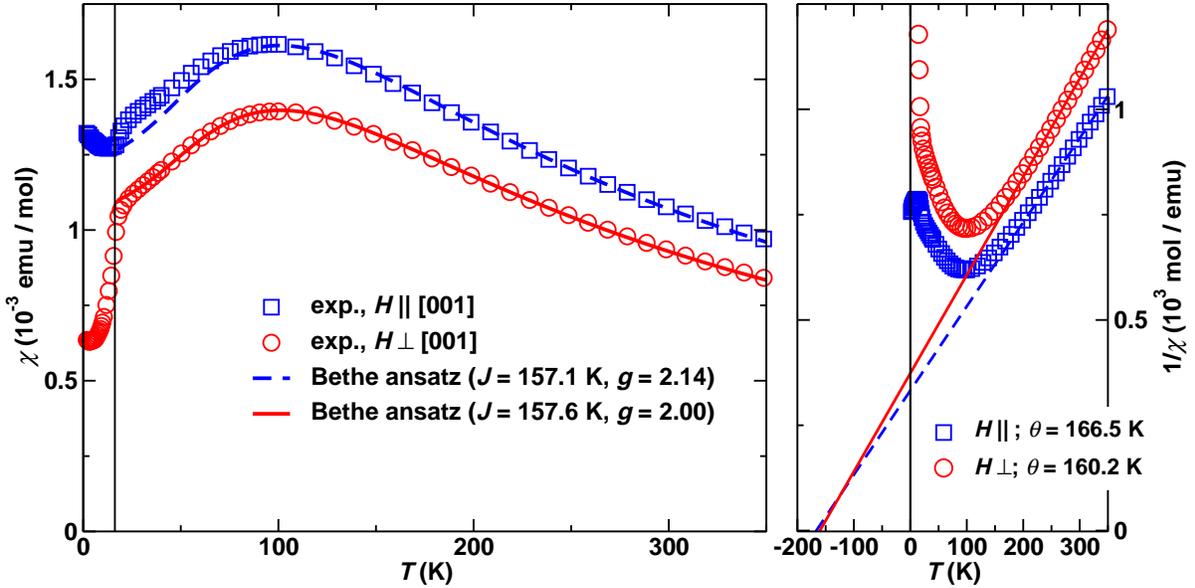}
\caption{\label{chi}
Left panel: magnetic susceptibility of CuSe$_2$O$_5$ as a function of
temperature. The magnetizing field is 10 kOe. For the graphic presentation, we
show only one of each five measured points. The Bethe ansatz fits are shown with
dashed and solid lines. Right panel: inverse magnetic susceptibility as a
function of temperature.  Curie-Weiss fits (for T~$>$~230~K) are shown with
lines. The temperature-independent contribution $\chi_0$ in the Curie-Weiss
fits was set to zero.}
\end{figure}

For CuSe$_2$O$_5$, the magnetic susceptibility curves for both field
orientations (figure~\ref{chi}, left panel) have a broad maximum at
$T_{\rm{max}}\approx101$~K and a finite value of $\chi$ at the lowest
temperature measured (1.8~K), indicating the low-dimensional behaviour and the
absence of a spin gap. The high-temperature parts of the curves obey the
Curie-Weiss law $\chi(T)=C/(T+\theta)$ (figure~\ref{chi}, right panel;
$T>220$~K, $H_{\parallel}$: $\theta=165$~K, $C=0.51$, $g=2.32$; $H_{\perp}$:
$\theta=170$~K, $C=0.43$, $g=2.15$).  The positive Curie-Weiss temperature
evidences that the dominating couplings in CuSe$_2$O$_5$ are antiferromagnetic. The
shape of the experimental curve reveals a close similarity to a spin-1/2
Heisenberg chain model. This model has an exact solution given by Bethe
ansatz~\cite{Bethe_Ansatz} and parametrized by Johnston et al.~\cite{johnston}.
We have fitted the experimental curves using the parametrized solution
(figure~\ref{chi}, left panel).

To account for the deviation of the fitted curves from experimental ones we
have fitted both curves independently\footnote{A simultaneous fit of both
curves implying the same $J$ value in the whole temperature range down to the
phase transition yields considerable deviations from experiment. The origin of
this deviation is discussed below in the text.} and varied the temperature
window. As a result, the magnetic susceptibility measured perpendicular to
needle-like crystallites can be perfectly fitted by a consistent set of
parameters ($J$~=~157.6~K, $g$~=~2.00) in the whole temperature range down to
the ordering temperature, while the fit to the susceptibility measured parallel
to the chains ($J$~=~157.1~K, $g$~=~2.14) shows deviations at low temperatures
(below $T_{\rm{max}}$). This difference likely originates from a slight
misalignment of microscopic plates in the needle-like crystallites. 

A phase transition is observed at 17~K for both orientations of the magnetizing
field. The nature of this magnetic transition can be understood by examination
of the low-temperature part of the curve (below the transition). The
interpretation is straightforward as soon as we account for (i) impurity
effects and (ii) effects of misalignment of the sample (relevant especially for
$\vec{H}_{\parallel}$, as shown above). Due to the high quality of samples, the
temperature region between the kink at 17~K down to 10~K is practically
unaffected by defects (no Curie tail). In this range, $\chi_{\perp}$ decreases
very slightly, while $\chi_{\parallel}$ drops distinctly on cooling, following
the theoretical result for ordered collinear
antiferromagnets~\cite{goodenough}. We assign the slight decrease of
$\chi_{\perp}$ (theory predicts it to be constant) to a small misalignment of
crystallites, in agreement with the deviations of the Bethe ansatz fit. The
small upturn in $\chi_{\parallel}$ at about 6~K (a zero susceptibility at zero
temperature follows from theory) is likely related to defects and paramagnetic
impurities.

\begin{figure}[htbp]
\includegraphics[angle=270,width=16cm]{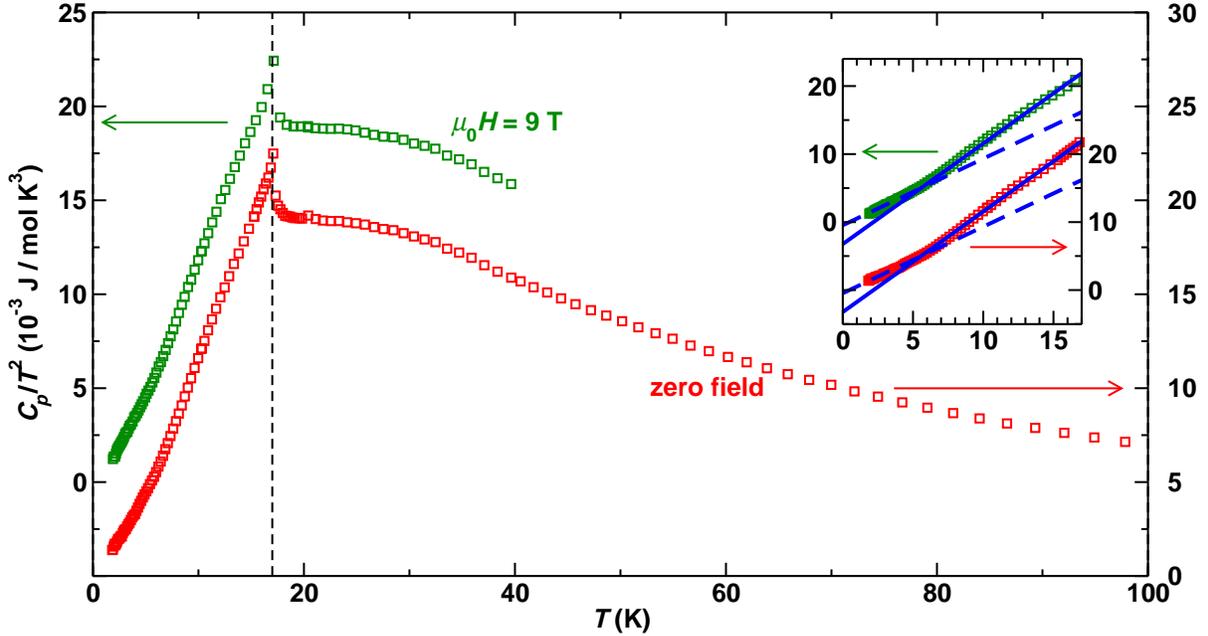}
\caption{\label{cp}
$C_p/T^2$ of CuSe$_2$O$_5$ as a function of temperature and magnetic field. The
N\'eel temperature is marked with a dashed line.  Inset (ordered phase region):
the $c_p{\sim}T^3$ behaviour predicted by theory is complicated by a clear kink
at 7~K.}
\end{figure}

To get additional information about the magnetic properties, we measured the
temperature dependence of the specific heat. The clear anomaly at~17~K (figure
\ref{cp}) and the linear behavior of $C/T^2(T)$ below this temperature are
typical for antiferromagnets~\cite{thermodynamics}. Thus, we interpret this as a
transition to an AFM ordered state ($T_{\rm{N}}$~=~17~K). Remarkably, the anomaly
does not shift nor decrease in amplitude in magnetic fields up to
$\mu_0H$~=~9~T. Prior to the analysis of the magnetic behaviour above
$T_{\rm{N}}$, the specific heat should be decomposed into the magnetic
contribution (which reflects the spectrum of magnetic excitations) and the
phonon contribution (the spectrum of lattice vibrations). This decomposition is
reliable only if the overlap of the two spectra (magnetic excitations and
phonons) is relatively small (see~\cite{BaCdVOPO42} for an example). As the
phonon contribution increases on temperature, the decomposition is possible for
systems with weak magnetic couplings ($J_{ij}<10$~K). As we obtained from our
susceptibility data, the energy scale of $J$ in CuSe$_2$O$_5$ is about 165~K.
Thus, for a 1D Heisenberg chain we expect the maximum of the magnetic specific
heat at 0.48$J$~\cite{johnston}, i.e. close to 80~K.  At this temperature, the
phonon contribution to the specific heat strongly dominates over the magnetic
contribution. As a result, the experimental curve has no visible features in the
vicinity of 80~K.  For systems with large couplings ($J_{ij}>10$~K), the most
accurate way to account for the phonon part is to measure an isostructural
non-magnetic reference system (see \cite{LiV2O4} for an example). In case of
CuSe$_2$O$_5$, it is not possible, as ZnSe$_2$O$_5$ has a different crystal
structure~\cite{ZnSe2O5}, and thus a different phonon spectrum. Therefore, the
specific heat data provide a clear evidence of an AFM ordering at 17~K but do not
allow an independent justification of the temperature scale for the leading
magnetic interactions.

The last remark concerns a pronounced kink at 7~K~(figure~\ref{cp}, inset), i.e.
the region of the ordered phase. The kink is stable at least up to
$\mu_0H$~=~9~T and thus not related to defects.  Intriguingly, a similar feature
has been observed for a related system Bi$_2$CuO$_4$ (figure~3
in~\cite{Bi2CuO4_cp}) favouring the intrinsic nature of the kink rather than a
sample dependent effect. For the magnetic contribution to the specific heat, such
features have been proposed to mark a dying out of high frequency spin wave
modes~\cite{atari}. To elucidate this unusual feature, further experimental
studies on CuSe$_2$O$_5$ and similar systems as well as a careful theoretical
analysis should be carried out.

%%%%%%%%%%%%%%%%%%%%%%%%%%%%%%%%%%%%%%%%%%%%%%%%%%%%%%%%%%%
%
% Microscopic model 
%
%%%%%%%%%%%%%%%%%%%%%%%%%%%%%%%%%%%%%%%%%%%%%%%%%%%%%%%%%%%

\subsection{Microscopic model}

\begin{figure}[htbp]
\includegraphics[width=12cm]{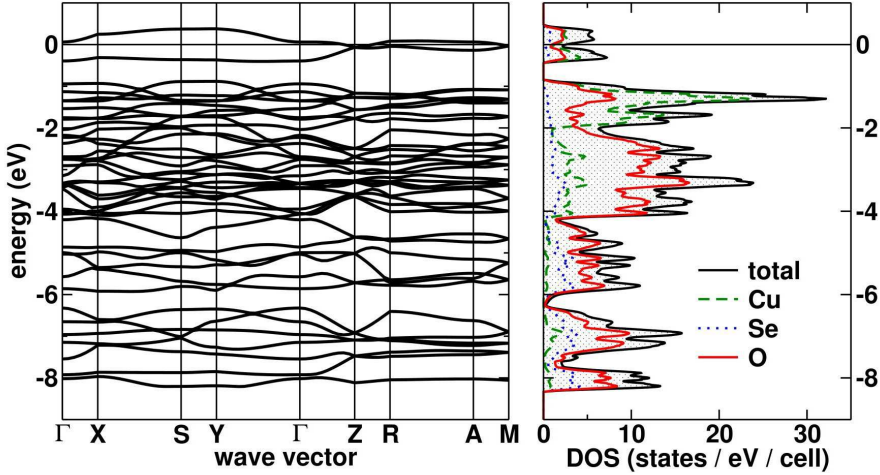}
\caption{\label{band_dos}
LDA band structure (left panel) and density of states (right panel) of
CuSe$_2$O$_5$. Notation of $k$-points: $\Gamma$=(000),
X=($\frac{\pi}{a}00$), S=($\frac{\pi}{a}\frac{\pi}{b}0$),
Y=($0\frac{\pi}{b}0$),
Z=(00$\frac{\pi}{c}$), R=($\frac{\pi}{a}0\frac{\pi}{c}$),
A=($\frac{\pi}{a}\frac{\pi}{b}\frac{\pi}{c}$), M=($0\frac{\pi}{b}\frac{\pi}{c}$).}
\end{figure}

We start in our microscopic analysis with band structure calculations performed
in the local density approximation (LDA). LDA yields a valence band of about 9
eV width formed mainly by Cu~$3d$, O~$2p$ and Se~$4p$ states
(figure~\ref{band_dos}, right panel). The well-separated double-peak at the
Fermi level $\varepsilon_{\rm{F}}$ contains two narrow, half-filled bands
(figure~\ref{band_dos}, left panel). The width of this antibonding band complex
(0.85~eV) is in between the widths of the same complex in Bi$_2$CuO$_4$
(1.05~eV~\cite{Bi2CuO4}) and Sr$_2$Cu(PO$_4$)$_2$ (0.65~eV~\cite{Sr2CuPO42}).
The LDA yields a metallic GS, contrary to the experimentally observed insulating
behavior.  This discrepancy is caused by the underestimate of strong on-site
Coulomb interactions of the Cu 3$d$ electrons. Nevertheless, LDA reliably
yields the relevant orbitals and dispersions. Thus, we have a closer look to
the band complex at $\varepsilon_{\rm{F}}$. The two bands, relevant for the
low-lying magnetic excitations, are related to the antibonding ${dp_{\sigma}}$
orbital of a CuO$_4$ plaquette, i.e. the antibonding combination of Cu
3$d_{x^2-y^2}$ and O 2$p_{\sigma}$ states (orbitals are denoted with respect to
the local coordinate system). The antibonding ${dp_{\sigma}}$ orbital is well
separated (${\Delta}E$~$\sim$~0.5~eV) from the lower lying Cu $3d$ and O $2p$
states. Thus, the most efficient way to describe the electronic structure is to
construct an effective one-band tight-binding (TB) model (one band per
plaquette), parametrized by a set of electron transfer integrals $t_{ij}$. The
correlation effects, insufficiently described by LDA and thus by the TB model,
are accounted for by adopting a corresponding Hubbard model mapped subsequently onto a
Heisenberg model (this mapping is valid for spin excitations in the
strongly correlated limit at half-filling, both well justified for undoped
cuprates with small magnetic exchange).

\begin{figure}[htbp]
\includegraphics[width=16cm]{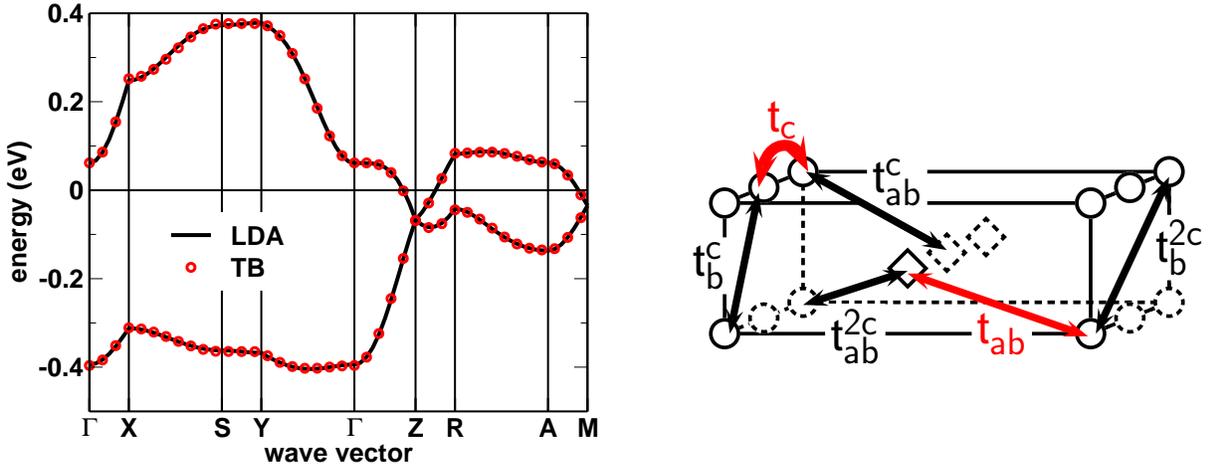}
\caption{\label{band_fit_ts}
Left panel: the tight-binding fit (circles) to the LDA band structure
(antibonding $dp_\sigma$ band, solid line) Right panel: the
superexchange paths for the leading transfer integrals.  The
projection of the structure is the same as in Fig.~\ref{str}.}
\end{figure}

Prior to numerical calculations, we compare the dispersions of the two
well separated bands at $\varepsilon_{\rm{F}}$ (figure~\ref{band_fit_ts}) to
dispersions of the corresponding antibonding $dp_{\sigma}$ complexes of
Sr$_2$Cu(PO$_4$)$_2$~(figure~2 in~\cite{Sr2CuPO42}) and Bi$_2$CuO$_4$
(figure 5 in~\cite{Bi2CuO4}). Here, a close similarity of
CuSe$_2$O$_5$ and Sr$_2$Cu(PO$_4$)$_2$ is revealed: both band structures have a
dominating dispersion along the chain direction (for CuSe$_2$O$_5$,
this is the $c$-axis in figure~\ref{str} and \mbox{$\Gamma$---Z} region in
figure~\ref{band_dos}) and a weaker dispersion in other directions,
unlike Bi$_2$CuO$_4$, where the dispersions along different directions
in the $k$-space are comparable, indicating a 3D behaviour.

For a quantitative analysis, we constructed an effective one-band TB Hamiltonian
and determined the set of transfer integrals $t_{ij}$ in order to get the best
least-squares fit to the two LDA bands crossing $\varepsilon_{\rm{F}}$.  As an
alternative approach, we used a Wannier functions (WF) technique which implies
the construction of WF for the Cu 3$d_{x^2-y^2}$ antibonding state, relevant for
the magnetism, and the calculation of the overlap of the WF. The results of the
latter method are affected by the overlap of the relevant 3$d_{x^2-y^2}$
antibonding state with other states.  Therefore, for perfectly separated bands
as in CuSe$_2$O$_5$, both methods should yield the same results within numerical
accuracy. The numerical evaluation supports this statement: the difference
between the transfer integrals obtained by the WF method and by the TB fit is
tiny and does not exceed 2~meV for individual $t_{ij}$ values (the mean value
for all $t_{ij}$ is 0.2~meV). This deviation can be considered as an error
margin for the mapping procedure. Thus, for isolated bands the WF method should
not be regarded more accurate than a direct TB fit, but rather as an independent
alternative procedure~\cite{salunke,salunke_comment,salunke_reply}. The
agreement of the results using the two independent mapping methods reflects the
applicability of an effective one-band approach.

The resulting set of the transfer integrals (table~\ref{exch}, first column)
yields perfect agreement with the LDA
bands~(figure~\ref{band_fit_ts}).\footnote{To check the results for consistency,
we have neglected all $t_{ij}$ smaller than 10 meV and repeated the fitting. The
difference of leading terms in both approaches did not exceed 10\%.} The hopping
paths corresponding to the leading terms are shown in figure~\ref{band_fit_ts}
(right panel).

\begin{figure}[htbp]
\includegraphics[width=16cm]{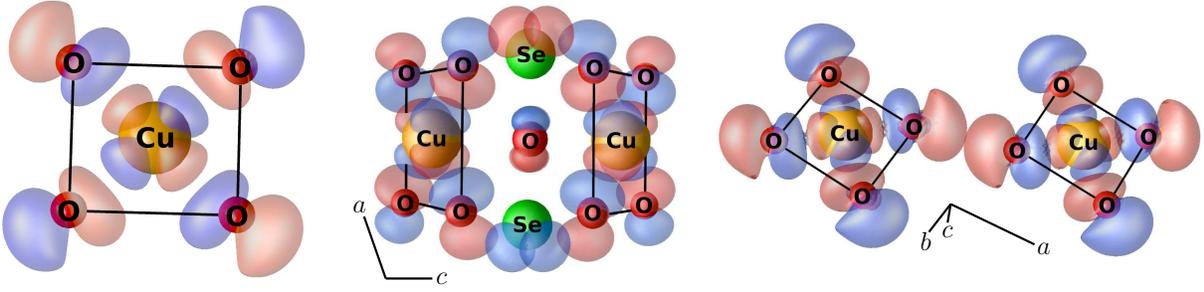}
\caption{\label{wf} Wannier functions for the Cu 3$d_{x^2-y^2}$
orbital. Colours represent the sign of a Wannier function. Left
panel: the Cu 3$d_{x^2-y^2}$ Wannier function plotted on top of a
CuO$_4$ plaquette, visualizing the antibonding combination of
Cu~3$d_{x^2-y^2}$ and O~$2p_{\sigma}$ states, relevant for the magnetism.
Central panel: the overlap of two Wannier functions centered on the
neighbouring Cu atoms (corresponds to the nearest neighbour
intra-chain coupling $t_c$). Note that the neighbouring plaquettes
are tilted which leads to a sizable $\pi$-overlap of the Wannier
functions, and hence O $2p$ wave functions of the neighbouring
plaquettes, allowing for a considerable ferromagnetic contribution to
the magnetic exchange. Right panel: the overlap of the Wannier
functions corresponding to the leading inter-chain coupling $t_{ab}$.}
\end{figure}

\begin{figure}[htbp]
\includegraphics[width=16cm]{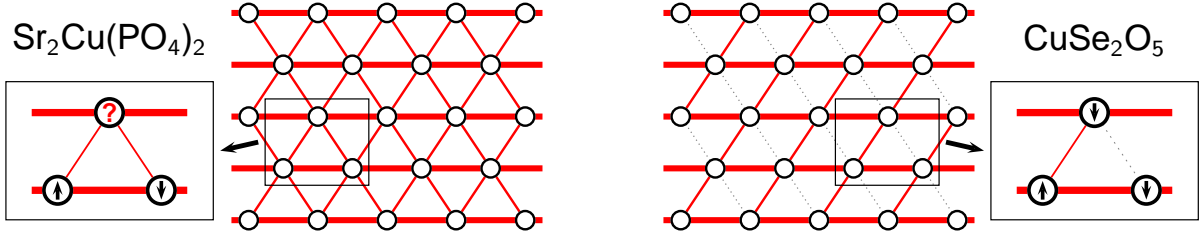}
\caption{\label{ic}Topology of inter-chain couplings in
Sr$_2$Cu(PO$_4$)$_2$ (left panel) and CuSe$_2$O$_5$ (right panel).
Red lines denote the AFM coupling. Bold red lines highlight the
chains. In Sr$_2$Cu(PO$_4$)$_2$, the intra-chain and the
two equivalent inter-chain couplings form an anisotropic triangular
lattice. In CuSe$_2$O$_5$, there is only one relevant inter-chain
coupling resulting in an anisotropic square lattice geometry of intra-chain and
inter-chain couplings. The former geometry leads to magnetic
frustration, while the latter is not frustrated (see insets).}
\end{figure}

We find that the leading couplings in CuSe$_2$O$_5$ are the NN intra-chain
coupling $t_c$~=~165~meV and one of the short inter-chain couplings
$t_{ab}$~$\approx$~45~meV. The corresponding Wannier functions are pictured in
figure~\ref{wf}. The value of the largest (NN intra-chain) coupling in
CuSe$_2$O$_5$ is slightly larger than the corresponding coupling in
Sr$_2$Cu(PO$_4$)$_2$ (135 meV~\cite{Sr2CuPO42}). The difference in the largest
inter-chain term is more pronounced: the size of the inter-chain coupling in
CuSe$_2$O$_5$ (45~meV) is considerably higher than in Sr$_2$Cu(PO$_4$)$_2$
(16~meV~\cite{Sr2CuPO42}). Even more important is the difference in the specific
coupling geometry --- whether it is constructive towards the long-range ordering
or not.  As we stated while comparing the crystal structures of CuSe$_2$O$_5$
and Sr$_2$Cu(PO$_4$)$_2$ (see section~\ref{sec_str}), in both systems there are
two short inter-chain coupling paths. The corresponding couplings are identical
(symmetry-related) in Sr$_2$Cu(PO$_4$)$_2$, but independent (and in fact,
considerably different) in CuSe$_2$O$_5$.  The TB analysis reveals that only one
($t_{ab}$) of the two NN inter-chain couplings is relevant for CuSe$_2$O$_5$
(table~\ref{exch}).  Consequently, the essential difference between the two systems can be
best understood in terms of the spin lattices that are formed by the strongest
intra-chain and inter-chain couplings, as depicted in figure~\ref{ic}. In
Sr$_2$Cu(PO$_4$)$_2$, three relevant couplings (the intra-chain NN coupling and
two identical inter-chain couplings) are arranged on an anisotropic triangular
lattice~(figure~\ref{ic}, left panel).  By switching off one of the inter-chain
couplings, the topology of the relevant couplings changes, and the system is
described by two couplings forming an anisotropic square
lattice~(figure~\ref{ic}, right panel). The main difference between the two
topologies is that in Sr$_2$Cu(PO$_4$)$_2$ the competition of relevant
couplings, which can not be simultaneously satisfied, leads to strong magnetic
frustration, while in CuSe$_2$O$_5$ the inter-chain couplings are not
frustrated.  The lifting of frustration in CuSe$_2$O$_5$ has a remarkable
influence on the physical properties as will be discussed below.

The calculated transfer integrals provide valuable information on the coupling
regime. To include the missing Coulomb interaction $U_{\rm{eff}}$, as described
in the section~\ref{method}, we can use the TB model to construct a Hubbard
model and map the latter onto a Heisenberg model to obtain the antiferromagnetic
(AFM) exchange from $J^{\rm{AFM}}_{ij}=4t_{ij}^2/U_{\rm{eff}}$. Using the same
representative $U_{\rm{eff}}$~=~4.5~eV as for
Sr$_2$Cu(PO$_4$)$_2$~\cite{Sr2CuPO42}, we obtain $J^{\rm{AFM}}_c$~=~285~K for
the NN intra-chain exchange and $J^{\rm{AFM}}_{ab}$~=~27~K for the largest
inter-chain exchange. Other couplings yield values of AFM exchange less than
1.5~K (table~\ref{exch}, second column) and will be neglected in further
discussion.

\begin{table}[htbp]
\caption{\label{exch}Leading transfer (first column) and exchange
integrals (last column) of CuSe$_2$O$_5$. The AFM exchange (second
column) is calculated via mapping the transfer integrals onto an
extended Hubbard ($U_{\rm{eff}}$ = 4.5 eV) and subsequently onto a
Heisenberg model. The total exchange is taken from \mbox{LSDA+$U$}
total energy calculations of supercells. The ferromagnetic exchange
$J^{\rm{FM}}_{ij}$ is evaluated as the difference between $J_{ij}$ and
$J^{\rm{AFM}}_{ij}$.} \begin{indented}
\item[] \begin{tabular}{@{}lllll}
		\br path & $t_{ij}$/meV & $J^{\rm{AFM}}_{ij}$/K &
		$J^{\rm{FM}}_{ij}$/K & $J_{ij}$/K \\ \mr
		 \textsl{X}$_c$              &166& 285 & $-$120& 165 \\ 
		 \textsl{X}$_{ab}$           & 51&  27 & $-$7  & 20  \\
		 \textsl{X}$^{2c}_{b}$       & 11& 1.5 & 0     & 1.5 \\ 
		 \textsl{X}$^c_{ab}$         & 10&   1 & 0     & 1   \\
		 \textsl{X}$^c_b$            & 10&   1 & 0     & $<$1\\ 
		 \textsl{X}$^{2c}_{ab}$      &  7& 0.5 & 0     & $<$1\\ \br
	\end{tabular}
\end{indented}	
\end{table}

The calculated leading magnetic exchange $J^{\rm{AFM}}_c$~=~285~K is
considerably larger than our estimate from the Bethe ansatz fit ($\approx$155~K)
based on experimental $\chi(T)$ data. Moreover, it is larger than the
corresponding exchange integral in Sr$_2$Cu(PO$_4$)$_2$ (187~K).  This
discrepancy originates from ferromagnetic (FM) contributions to the total
magnetic exchange, which are neglected in the mapping procedure. For the NN
exchange $J_c$, we expect a considerable FM contribution originating from the
overlap of O $2p$ wave functions of neighbouring plaquettes. Due to a dihedral
angle $\phi$~=~64$^{\circ}$ between the neighbouring plaquettes, this overlap
has a sizable $\pi$ contribution~(this can be seen in the WF in figure~\ref{wf},
central panel) leading to a Hund's rule (FM) coupling. For the leading
inter-chain coupling, the FM contribution is expected to be small due to a
predominantly $\sigma$ overlap of O $2p$ wave functions~(figure~\ref{wf}, right
panel).

To get a numerical estimate for the FM contribution, we perform total
energy calculations for various spin patterns of magnetic supercells
using the \mbox{LSDA+$U$} method. The method is rather sensitive to
the $U_d$ value. As $U_d$~=~6.5~eV yields agreement between the
calculated and experimentally measured exchange integrals of the well
studied La$_2$CuO$_4$ and CuGeO$_3$, we adopted this value in the
calculations for CuSe$_2$O$_5$.\footnote{It is worth to note, that the
$U_d$ parameter is not universal and depends on a calculational scheme
and consequently on the basis set implemented in a code. Thus, different
$U_d$ values adopted in this work for CuSe$_2$O$_5$ ($U_d$~=~6.5~eV, the
code fplo version 7.00-27) and for Sr$_2$Cu(PO$_4$)$_2$
($U_d$~=~8.0~eV~\cite{Sr2CuPO42}, the code fplo version 5.00-18)
originate from the different basis used in the codes.} The supercell
method has limitations set by numerical accuracy for the small
exchange integrals ($J_{ij}<1$~K) and the size of the required supercells.
In our case, we constructed supercells and spin patterns that yield
all exchange couplings which were found to be
relevant from the TB analysis. The resulting total energies
are mapped onto a Heisenberg model, which is parameterized by the
total exchange integrals (table~\ref{exch}, last column) containing both AFM
and FM contributions. Thus, by subtracting the AFM part
$J^{\rm{AFM}}_{ij}$ from the total exchange $J_{ij}$, the FM contribution
$J^{\rm{FM}}_{ij}$ can be estimated (table~\ref{exch}, fourth column).

In general, \mbox{LSDA+$U$} calculations yield a reliable estimate for exchange
integrals~\cite{Sr2CuPO42,K2CuP2O7,Bi2CuO4,BaCdVOPO42}.  This
reliability holds for CuSe$_2$O$_5$: we obtain $J_c$~=~165~K and
$J_{ab}$~=~20~K, in almost perfect agreement with the estimates from magnetic
susceptibility. In accordance with our expectations, $J^{\rm{FM}}_c$~=~$-120$~K
has a considerable contribution to the total exchange $J_c$, while
$J^{\rm{FM}}_{ab}$~=~$-7$~K yields a smaller correction to the $J_{ab}$ value.
We should note that the large ferromagnetic contribution
$J^{\rm{FM}}_c$~=~$-120$~K may originate, in addition to the mentioned
$\pi$-overlap of O $2p$ wave functions, also from a destructive interference of
coupling paths~\cite{exchange_interference_paths} or by a strong coupling to
ligands~\cite{exchange_side_groups}. Which of these mechanisms plays a leading
role in CuSe$_2$O$_5$ is an open question. This issue is, however, beyond the
scope of the present paper and needs further theoretical investigation.

Though the calculated $J_c$ value (165~K) is very close to the
estimate from the Bethe ansatz (157~K), we decided to check the
exchange integrals for consistency by performing additional
calculations for $U_d$~=~6.0~eV and $U_d$~=~7.0~eV. Besides the
expected change of exchange integrals (0.5~eV increase of the
$U_d$ results in about 20\% decrease of $J_{ij}$ and vice versa), we found that
the ratio $\alpha$~${\equiv}$~$J_{ab}/J_{c}$ of the leading exchange
integrals ($\alpha$~=~0.121 for $U_d$~=~6.0~eV, $\alpha$~=~0.129 for
$U_d$~=~6.5~eV and $\alpha$~=~0.136 for $U_d$~=~7.0~eV) is rather
stable with respect to the $U_d$ value.

Thus, the consideration of the FM contribution yielded a valuable
improvement of the energy scale comparing to the AFM exchange values
$J^{\rm{AFM}}_c$ and $J^{\rm{AFM}}_{ab}$, but the ratio $\alpha$ of the two
couplings, that is the most relevant for the magnetic ground state,
stays almost unchanged. Moreover, this ratio is stable with respect to
the model parameters $U_{\rm{eff}}$ and $U_d$, leading to a very reliable
physical picture: CuSe$_2$O$_5$ can be described as a quasi-1D system
with AFM chains  characterized by an NN intra-chain exchange of 165~K.  Each chain
is coupled to two neighbouring chains by the non-frustrated
inter-chain exchange of 20~K (one order of magnitude smaller that the
intra-chain coupling). 

%%%%%%%%%%%%%%%%%%%%%%%%%%%%%%%%%%%%%%%%%%%%%%%%%%%%%%%%%%%
%
% Simulations
%
%%%%%%%%%%%%%%%%%%%%%%%%%%%%%%%%%%%%%%%%%%%%%%%%%%%%%%%%%%%

\subsection{Simulations}
\begin{figure}[htbp]
\includegraphics[angle=270,width=16cm]{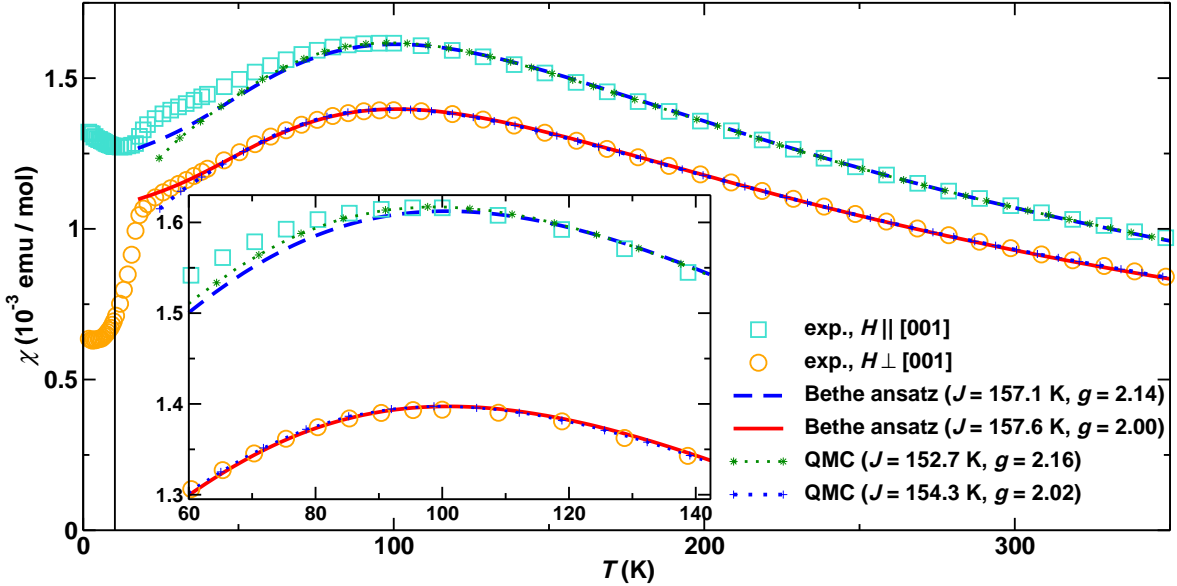}
\caption{\label{chi_fit}
Comparison of Bethe ansatz and quantum Monte Carlo (QMC) fits to the
experimental magnetic susceptibility. The temperature-independent
contribution $\chi_0$ in both fits was set to zero.}
\end{figure}

Turning back to the discussion about the ordering temperature
(section~\ref{intro}), it is reasonable to point out the advantages of
CuSe$_2$O$_5$ as a model system. First, we have evidence from both theory and
experiment that the system is mainly 1D.  Secondly, the microscopic analysis
revealed that the NNN intra-chain coupling is practically absent leading to a
valuable simplification for a theoretical analysis. Finally, there is only one
relevant inter-chain coupling. The fact that this latter coupling is not
frustrated allows to use the powerful quantum Monte Carlo (QMC) method for a
simulation of thermodynamical data with a subsequent comparison to the
experimentally measured curves. The results of the simulations are given in
figure~\ref{chi_fit} in comparison with the Bethe ansatz fits (where the
inter-chain coupling is neglected).  Obviously, the inclusion of the
inter-chain coupling yields only a tiny improvement with respect to the Bethe
ansatz fits.  This fact demonstrates $a$ $posteriori$ the importance of a
microscopic model for systems like CuSe$_2$O$_5$: apart from the microscopic
modelling, there is no reliable way to account for the small inter-chain
coupling directly from measurements of the paramagnetic susceptibility.

%%%%%%%%%%%%%%%%%%%%%%%%%%%%%%%%%%%%%%%%%%%%%%%%%%%%%%%%%%%
%
% Estimation of the Neel temperature
%
%%%%%%%%%%%%%%%%%%%%%%%%%%%%%%%%%%%%%%%%%%%%%%%%%%%%%%%%%%%

\subsection{Estimation of the N\'eel temperature}
To benefit from the unique combination of the simple microscopic
picture and the experimentally well-determined AFM ordering
temperature~$T_{\rm{N}}$, we
make an attempt to estimate $T_{\rm{N}}$ from an available simplified theory using the calculated
exchange integrals. Still, there are three problems, intrinsic for
quasi-1D systems, to be accounted for. The first is the spatial
anisotropy of exchange couplings present in a system.  This problem is
resolved in CuSe$_2$O$_5$ only partially. On one hand, there is only
one relevant inter-chain coupling for every pair of neighbouring
chains, but on the other hand, it couples a certain chain with only
two of four neighbouring chains. Thus, the couplings to the other two
chains are considerably smaller, resulting in the (spatial) exchange
anisotropy. The second problem is the anisotropy in the spin space. In
our microscopic approach, we used the isotropic Heisenberg model where
this anisotropy is neglected. Although we observe a remarkable agreement
between the microscopic model and the macroscopic behaviour, the spin
anisotropy is present, as evidenced for instance by the strong
dependence of the $g$-factor on the orientation of a magnetizing field
(figure~\ref{chi}). In a common sense approach, the (spatial) exchange
anisotropy is expected to lower $T_{\rm{N}}$, while the spin anisotropy
raises it. In each system, these two effects are balanced. Attempts to
find a suitable description for this balance were made in a number of
advanced theoretical studies based on a mean-field
formalism~\cite{tN_scalapino,tN_schulz},~\cite{tN_irkhin}\nocite{tN_bocquet,tN_yasuda}--\cite{tN_todo}.
Still, the problem seems not to be resolved, since a considerable disagreement
remains between the numerical results
yielded by different theories (none of which is generally accepted)
and, even more important, due to the third problem --- the problem of
inter-chain magnetic frustration, which has not been addressed, so far. 

Here, we make an empirical attempt to estimate how the frustration
influences the magnetic ordering. For that purpose, we compare several
quasi-1D magnetic compounds in a systematic way. The two well studied
quasi-1D cuprates Sr$_2$CuO$_3$ and Ca$_2$CuO$_3$ are
commonly referred as model systems in most theoretical studies
regarding the $T_{\rm{N}}$ problem. We should note that these two systems are
essentially different from CuSe$_2$O$_5$ due to the presence of
corner-sharing chains of CuO$_4$ plaquettes, which results in one
order of magnitude larger NN coupling. In addition, the
NNN coupling is not negligible~\cite{Sr_Ca2CuO3}.
Nevertheless, they are referred here for the sake of completeness. In
both, Sr$_2$CuO$_3$ and Ca$_2$CuO$_3$, one of the relevant inter-chain couplings
is frustrated. To calculate $T_{\rm{N}}$, we use formulas given by
Schulz~\cite{tN_schulz}, Irkhin and Katanin~\cite{tN_irkhin} and
Yasuda et al.~\cite{tN_yasuda}. The calculated $T_{\rm{N}}$ are given in the
three last columns of (table~\ref{compar}). Disregarding the method
used, the calculated $T_{\rm{N}}$ considerably overestimate the experimental
values for Sr$_2$CuO$_3$ and Ca$_2$CuO$_3$ (table~\ref{compar}, fifth
column).

\begin{table}[htbp]
\caption{\label{compar}
Exchange integrals (columns 2 and 3) together with experimental (column 5) and
theoretically calculated (columns 7--10) ordering temperatures $T_{\rm{N}}$ for
quasi-one-dimensional cuprates (3D magnet Bi$_2$CuO$_4$ has a similar structural
motive and was added for completeness). The ordering temperatures $T_{\rm{N}}$
were calculated using formulas from
references~\cite{tN_schulz,tN_irkhin,tN_yasuda} and the exchange integrals from
columns 2 (nearest neighbour intra-chain coupling $J_1$) and 3 (leading
inter-chain coupling $J_{\perp}$). Note that the first four systems are
frustrated due to inter-chain couplings, while in the last two the inter-chain
couplings are not frustrated.}
\begin{indented}
\item[]	\begin{tabular}{@{}llllllllll}
		\br compound & $J_1$/K & $J_{\perp}$/K & Ref. & $T_{\rm{N}}$ exp./K & Ref. & \multicolumn{4}{c}{$T_{\rm{N}}$ calc./K} \\ 
		& & & & & & ~\cite{tN_schulz} & \cite{tN_irkhin} & ~\cite{tN_yasuda} & \cite{Bi2CuO4}\\ \mr
		Sr$_2$CuO$_3$        & 2200 & 9    & \cite{Sr_Ca2CuO3} & 5 &\cite{Sr_Ca2CuO3_tN} & 28  & 22 & 21 \\
		Ca$_2$CuO$_3$        & 1850 & 42   & \cite{Sr_Ca2CuO3} & 9 &\cite{Sr_Ca2CuO3_tN} & 115 & 91 & 85 \\
		Sr$_2$Cu(PO$_4$)$_2$ & 187  & 3    & \cite{Sr2CuPO42}  & 0.085 &\cite{Sr2CuPO42_Tn} & 8.5 & 6.7 & 6.3 \\
		K$_2$CuP$_2$O$_7$    & 196  & 0.25 & \cite{K2CuP2O7} & $<2$ & \cite{K2CuP2O7} & 0.9 & 0.7 & 0.6 \\
		CuSe$_2$O$_5$        & 165  & 20   &  & 17  &  & 23  &	18 & 17 \\
		Bi$_2$CuO$_4$        & 10   & 6    & \cite{Bi2CuO4} & 42 & \cite{bicu_tN} & & & & 47 \\ 
		\br
	\end{tabular}
\end{indented}	
\end{table}

A theoretical approach is expected to work better for systems with more
pronounced 1D nature --- e.g. Sr$_2$Cu(PO$_4$)$_2$ and K$_2$CuP$_2$O$_7$. The
structural peculiarities of these systems were discussed in
section~\ref{sec_str}. The main issue here is the frustration caused by the
leading inter-chain coupling. For Sr$_2$Cu(PO$_4$)$_2$, theory predicts an
ordering temperature $T_{\rm{N}}$ two orders of magnitude larger than the
experimentally observed value. (Unfortunately, experimental low-temperature data
are not available for K$_2$CuP$_2$O$_7$.) This huge discrepancy is in sharp
contrast with the situation for the 3D magnet Bi$_2$CuO$_4$, for which the
theoretical estimate coincides with the experimental value within the error bars
(table~\ref{compar}, last row).\footnote{For the theoretical estimation of
$T_{\rm{N}}$, the formula 7 from~\cite{Bi2CuO4} was used.}

In CuSe$_2$O$_5$, each chain is strongly coupled only with two of four
neighbouring chains (unlike Sr$_2$CuO$_3$, Ca$_2$CuO$_3$, Sr$_2$Cu(PO$_4$)$_2$
and K$_2$CuP$_2$O$_7$ with coupling to four neighbouring chains). There is no
unique way to take this feature into account. A simple approximation is to take
the arithmetic average, which yields an effective inter-chain coupling value
$J_{\perp}=J_{ab}/2$.\footnote{Alternatively, the geometrical averaging can be
used. This approach yields a correct limit with respect to the Mermin-Wagner
theorem (zero ordering temperature for 1D and 2D systems). Then, calculational
schemes from~\cite{tN_schulz},~\cite{tN_irkhin} and~\cite{tN_yasuda} yield
$T_{\rm{N}}$ values of 14~K, 11~K and 10~K, respectively.} Using this value,
theory yields a perfect agreement with experimental value (table~\ref{compar},
fifth row).

Obviously, the existing models describe the magnetic ordering in CuSe$_2$O$_5$
much better than in Sr$_2$CuO$_3$ and Ca$_2$CuO$_3$, and especially in
Sr$_2$Cu(PO$_4$)$_2$. Despite our crude way of accounting for spatial exchange
anisotropy and the neglect of spin anisotropy, the theoretical estimate of
$T_{\rm{N}}$ for CuSe$_2$O$_5$ is in surprisingly good agreement with the
experimental value. Though in general the magnetic ordering is affected by the
spin anisotropy, CuSe$_2$O$_5$ yields empirical evidence that for systems with a
small spin anisotropy the isotropic model provides a rather accurate estimate of
$T_{\rm{N}}$.  Thus, it is unlikely that the disagreement between theoretical
and experimentally observed $T_{\rm{N}}$ values for Sr$_2$Cu(PO$_4$)$_2$
originates from the neglect of spin anisotropy effects.

Finally, only the magnetic frustration is left to be a possible reason for a
huge discrepancy between theory and experiment. Our analysis reveals that
frustrated inter-chain couplings play a crucial role for the magnetic
ordering. This fact explains why theoretical schemes fail to predict
$T_{\rm{N}}$ for frustrated systems. 

To illustrate the influence of frustration, we use a simple formula from the
spin wave theory in a random phase approximation, which connects N\'eel
temperatures for two compounds A and B with the values of exchange integrals:
$T^{A}_{N}$/$T^{B}_{N}$$\approx\sqrt{J^{\rm{A}}_1J^{\rm{A}}_{\perp}}/\sqrt{J^{\rm{B}}_1J^{\rm{B}}_{\perp}}$~\cite{Sr_Ca2CuO3}.
Using the values of exchange integrals for Sr$_2$Cu(PO$_4$)$_2$ and
CuSe$_2$O$_5$ and the experimental N\'eel temperature for Sr$_2$Cu(PO$_4$)$_2$
(table~\ref{compar}), we obtain $T_{\rm{N}}$~$\approx$~0.146~K for
CuSe$_2$O$_5$, almost 120 times smaller than the experimental value.

In the existing theoretical approaches, a parameter controlling the frustration
caused by inter-chain couplings is missing. Therefore, new theories which would
treat magnetic frustration as one of the key issues for the magnetic ordering, are
needed. On the other hand, there is a lack of information from the experimental
side, resulting in a very limited number of systems that challenge the
theoretical predictions.  Besides CuSe$_2$O$_5$, an almost perfect model system,
synthesis and investigation of new systems with similar crystal chemistry are
highly desirable.

%%%%%%%%%%%%%%%%%%%%%%%%%%%%%%%%%%%%%%%%%%%%%%%%%%%%%%%%%%%
%
% Summary
%
%%%%%%%%%%%%%%%%%%%%%%%%%%%%%%%%%%%%%%%%%%%%%%%%%%%%%%%%%%%

\section{\label{summ}Summary and outlook}
The class of quasi-1D magnets attracts much attention as a field of search for
prominent models and a playground for modern theories.  Recently, by studying
the magnetic properties of Cu$^{2+}$ phosphates, several systems of this class
were found to exhibit the physics of a Heisenberg chain model. In these
materials, the remarkable one-dimensionality and the absence of long-range
intra-chain interactions are ruled by a unique arrangement of magnetically
active CuO$_4$ plaquettes: they form edge-sharing chains where every second
plaquette is cut out. The chains are well separated by alkaline or alkaline
earth cations (K, Sr). The magnetic susceptibility of these systems is
perfectly described by the Bethe ansatz, which provides an exact solution for
the nearest neighbour (NN) antiferromagnetic spin-1/2 Heisenberg chain. At the
same time, the ordering temperature $T_{\rm{N}}$ of the systems reveals a fundamental
disagreement between theory and experiment. Unfortunately, the range of
available experimental studies of these systems is rather limited, as the
materials are presently available only as powders.

Therefore, we have synthesized CuSe$_2$O$_5$ --- a system implying a similar,
isolated arrangement of neighbouring CuO$_4$ plaquettes (but tilted with respect
to each other, unlike Cu$^{2+}$ phosphates) and allowing for a growth of high
quality single crystals.  Thermodynamic measurements reveal a quasi-1D behaviour
with a leading antiferromagnetic coupling of about 160~K (obtained from the
Bethe ansatz fit for the magnetic susceptibility).  The system orders
antiferromagnetically at 17~K, as evidenced by magnetic susceptibility and
specific heat data.  A microscopic analysis based on the results of DFT
calculations reveals that CuSe$_2$O$_5$ can be described in good approximation
by only two relevant exchange integrals: NN intra-chain ($J_c=165~K$) and the
leading inter-chain coupling ($J_{ab}=20~K$). The theoretical estimate of the
ordering temperature
$T_{\rm{N}}$ is in perfect agreement with experimental value.  This remarkable
agreement is in sharp contrast with a huge overestimate of $T_{\rm{N}}$ for
Cu$^{2+}$ phosphates yielded by a formal application of the same theory. To
reveal the origin of this difference on empirical grounds we analyzed
systematically the factors affecting $T_{\rm{N}}$. Beyond the influence of the
spatial exchange anisotropy and the spin anisotropy, we emphasize the role of
the magnetic frustration due to equivalent inter-chain interactions in the
latter compounds.  Comparing theoretical and experimental data for related
systems, we show that inter-chain frustrations have a crucial influence on
$T_{\rm{N}}$ and are likely the main cause for the failure of any theory which
ignores them. 

For an outlook, we propose further experimental studies (for instance, Raman
spectroscopy and inelastic neutron scattering) in order to benefit from the
availability of single crystals of CuSe$_2$O$_5$.  Especially, additional
experimental data are required in order to understand the nature of the specific
heat anomaly at 7~K.  Secondly, we hope that our work will inspire a directed
search for new quasi-1D model systems. Last but not the least, we want to
stimulate the
development of more sophisticated theories for the estimation of $T_{\rm{N}}$.
In particular, such theories should explicitly take into account the magnetic
frustration arising from complex inter-chain interactions. The well understood
system CuSe$_2$O$_5$ could give valuable support for these theories.

\section*{Acknowledgements} 

We are grateful to P~Scheppan for her help with the EDXS analysis and to
R~Cardoso-Gil and J~A~Mydosh for critical reading of the manuscript and
valuable suggestions. The investigation was supported by the GIF
(I-811-237.14103) and by the Emmy Noether program of the DFG.

\end{document}